\documentclass{article}

\usepackage{arxiv}
\usepackage{xurl}
\usepackage{amsmath}
\usepackage[utf8]{inputenc} % allow utf-8 input
\usepackage[T1]{fontenc}    % use 8-bit T1 fonts
\usepackage{hyperref}       % hyperlinks
\usepackage{url}            % simple URL typesetting
\usepackage{booktabs}       % professional-quality tables
\usepackage{amsfonts}       % blackboard math symbols
\usepackage{nicefrac}       % compact symbols for 1/2, etc.
\usepackage{microtype}      % microtypography
\usepackage{lipsum}	
\usepackage{graphicx}
\usepackage{natbib}
\usepackage{doi}

\usepackage{xcolor}
\usepackage{graphicx}
\usepackage{booktabs}
\usepackage{xspace}
\usepackage{cleveref}
\usepackage{fontawesome5}
\usepackage{tikz}
\usetikzlibrary{arrows.meta}
\usepackage{minted}
\usepackage{listings}
\usepackage{inconsolata}
\usepackage{wrapfig}
\usepackage{hyperref}

\definecolor{MColor}{HTML}{D55E00}
\definecolor{AColor}{HTML}{009E73}
\definecolor{DColor}{HTML}{0072B2}

\newcommand{\madlibs}{%
  \textbf{%
    \textcolor{MColor}{M}%
    \kern-0.15em%
    \raisebox{1.5pt}{\rotatebox[origin=c]{10}{\textcolor{AColor}{A}}}%
    \kern-0.2em%
    \raisebox{-2.25pt}{\rotatebox[origin=c]{-7}{\textcolor{DColor}{D}}}%
    \kern-0.15em%
    \raisebox{0.5pt}{\rotatebox[origin=c]{10}{l}}%
    i\kern-0.05em%
    \raisebox{-2.25pt}{\rotatebox[origin=c]{-7}{b}}%
    \kern-0.05em%
    s%
  }\xspace%
}

\definecolor{terminalbg}{HTML}{1E1E1E}
\definecolor{terminalfg}{HTML}{D4D4D4}
\definecolor{terminalborder}{HTML}{555555}
\definecolor{terminalcmd}{HTML}{4EC9B0}
\definecolor{terminalkey}{HTML}{9CDCFE}
\definecolor{terminalvalue}{HTML}{CE9178}

\lstdefinelanguage{HydraShell}{
    sensitive=true,
    alsoletter={._},
    morekeywords=[1]{python3},
    morekeywords=[2]{
        prompt_strategy,
        generator,
        analysis_type,
        run.n_prompts,
        run.output_dir,
        hydra.run.dir
    },
    morekeywords=[3]{
        dependency_graph,
        acestep,
        mir,
        output_dir,
        run_dir
    }
}

\lstdefinestyle{terminal}{
    language=HydraShell,
    basicstyle=\ttfamily\footnotesize\color{terminalfg},
    backgroundcolor=\color{terminalbg},
    keywordstyle=[1]\color{terminalcmd}\bfseries,
    keywordstyle=[2]\color{terminalkey},
    keywordstyle=[3]\color{terminalvalue},
    frame=single,
    framerule=0pt,
    framesep=5pt,
    columns=fullflexible,
    keepspaces=true,
    showstringspaces=false,
    breaklines=true,
    aboveskip=7pt,
    belowskip=7pt
}

\makeatletter
\let\arxiv@maketitle\@maketitle
\def\@maketitle{\def\@makefnmark{\hbox{$^{\@thefnmark}$}}\arxiv@maketitle}
\newcommand{\samethanks}[1][\value{footnote}]{\footnotemark[#1]}
\makeatother

\title{\madlibs: Analyzing the Distribution of Generative Music}

\author{Matthew Keating\thanks{Equal contribution}, Noah Schaffer\samethanks, Michael Casey, 
\& Nikhil Singh \\
Dartmouth College\\
\texttt{\{matthew.b.keating.gr,noah.schaffer.gr,michael.a.casey,nikhil.u.singh\}@dartmouth.edu} 
}

\date{}

\renewcommand{\headeright}{Keating et al.}
\renewcommand{\undertitle}{}
\renewcommand{\shorttitle}{\madlibs}

\hypersetup{
pdftitle={MADlibs: Analyzing the Distribution of Generative Music},
pdfauthor={Matthew Keating, Noah Schaffer, Michael Casey, Nikhil Singh},
pdfkeywords={Music generation, NLP for music, evaluation framework},
}
\begin{document}
\maketitle
\vspace{-0.4in} 

\begin{figure*}[!htb]
    \centering
    \includegraphics[width=\linewidth]{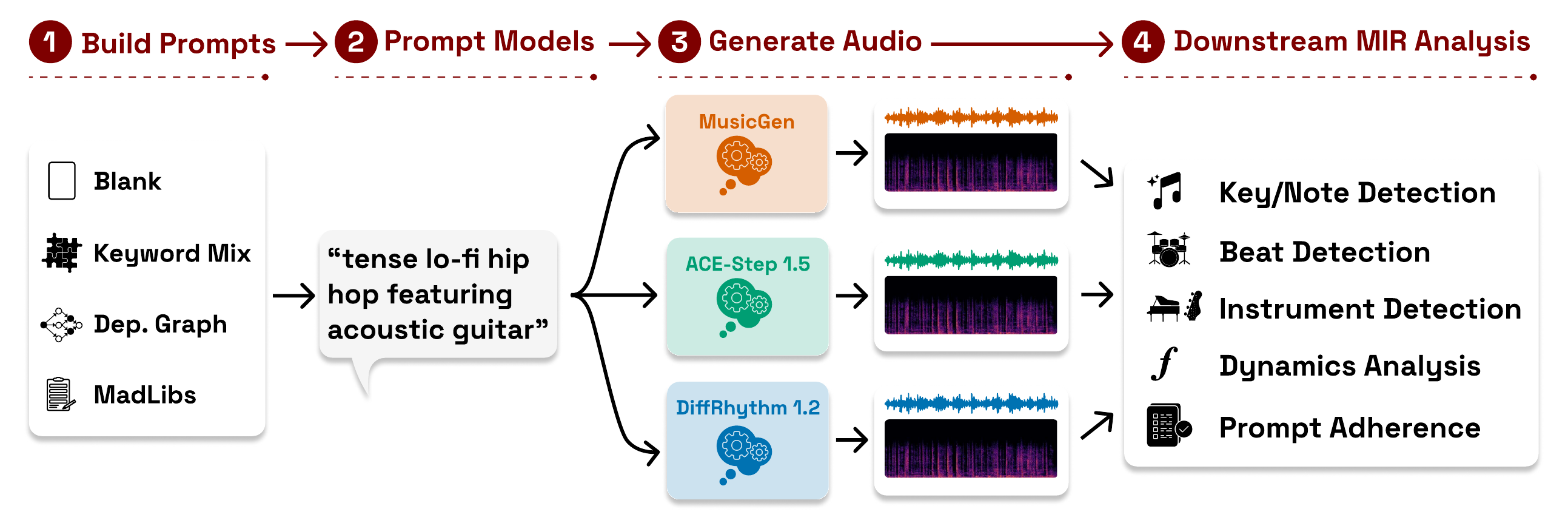}
    \caption{End-to-end \madlibs pipeline. Using five different prompting strategies, we generate $10,000$ prompts and use these as input to three text-to-music models. We then perform downstream MIR analysis on the output of all models. }
    \label{fig:banner}
\end{figure*}

\begin{abstract}
	What music can a text-to-music model actually make? The answer to this question has consequences for individual creators, communities, and musical culture more broadly. We introduce \madlibs, a framework to measure the \textit{expressive support} of state-of-the-art music generators: the subset of musical behaviors they will actually produce across the space of prompts. We first develop strategies for generating musical prompts using natural language processing methods and existing caption corpora. Then, we use a set of MIR techniques to study the corresponding output spaces in terms of interpretable and qualitatively meaningful elements like key, beat, tempo, instrumentation, and dynamics. We find that text-to-music models are not interchangeable instruments for the same prompt space. Rather, each model has learned idiosyncratic properties that constrain the regime of music reachable via prompting it, for example, defaulting to certain tempi, failing to produce silences, or oversampling canonical band arrangements. Because users interact with these systems one prompt at a time, these biases are often invisible in practice yet cumulatively shape what AI-assisted music sounds like, risking the production of a new auditory monoculture hidden behind the promise of universal describability. We thus urge the MIR community to further study and responsibly steer population-level properties of music generation models.
\end{abstract}

% keywords can be removed
% \keywords{Music generation \and NLP for music \and evaluation framework}

\section{Introduction}\label{sec:introduction}

The premise underlying text-to-music generation is that if you can describe it, you can create it. Though this assumption has driven progress in training generation models and even deploying them~\citep{copet2023simple, agostinelli2023musiclm}, it is a difficult one to test since it requires us to measure what precisely lies within the space of music such models can produce from prompts. This question is entangled with another important one: do these models preserve the diversity of their training data, or (inadvertently) concentrate the output probability mass on a narrower, mode-like subset of musical behaviors? If generative models strongly preferentially produce certain characteristics, they may implicitly constrain the creative space available to users, shaping musical output and ultimately culture in ways that are difficult to detect through casual use. Moreover, different model architectures and training procedures may induce different biases over this output space, leading to divergent notions of what is producible by prompting.

In this work, we introduce \madlibs, a framework for probing the output distribution of text-to-music models by sampling broadly over the prompt space and analyzing the resulting audio with music information retrieval techniques. We study three widely-used generative music models: \textbf{\textcolor{MColor}{M}}usicGen, \textbf{\textcolor{AColor}{A}}CE-Step, and \textbf{\textcolor{DColor}{D}}iffRhythm. Rather than evaluate models on prompt fidelity, we ask a more basic question: given a space of possible prompts, what kinds of music do these models actually produce? We define this as the \textit{expressive support} of a music generation model. The name \madlibs comes from the popular word game \textit{Mad Libs}, where a player fills out a story by randomly selecting words that fit a part of speech.

\madlibs is designed to evaluate the output of text-to-audio models when prompts are sampled from uniform-like distributions over the space of all reasonable prompts. To perfectly model this output distribution, one would need to be able to generate outputs for any possible prompt that could be handled by a generative music model. However, the English language has roughly a million words \citep{Merriam-Webster}, so this problem quickly becomes computationally intractable. Additionally, the underlying training data for most music generation models is not publicly available, so we cannot directly sample prompts from this distribution. We instead solve this feasibility challenge through a set of pseudo-random text sampling strategies.

Overall, this work contributes:
\begin{itemize}
    \item \textbf{Methods for producing music generation prompts} to probe the output distribution of large generative music models
    \item An \textbf{analysis suite} that combines classical MIR approaches with contemporary MIR models to produce interpretable summaries of this distribution (which we term the \textit{expressive support} of a given model)
    \item \textbf{Empirical evidence} that the expressive support of music generation models is limited in predictable ways relative to human-made music, and is more model-constrained than prompt-steerable
\end{itemize}

\section{Related Work}\label{sec:related}

Recent years have seen an influx of work in developing text-prompted music synthesis models. Despite differences in transformer-based \citep{copet2023simple, agostinelli2023musiclm, garcia2023vampnet}, diffusion-based \citep{chen2024musicldm} and diffusion-transformer \citep{evans2024long, gong2025ace,gong2026ace, ning2025diffrhythm,jiang2025diffrhythm} architectures, all text-conditioned generative music models share the same basic premise: anything you can describe well enough in text can be synthesized by the model.

However, there is growing evidence that there are systematic failure modes in modern generative models. Several works on language models and multimodal models have shown that one can adversarially craft semantically diverse input that produces the same output in a generative model \citep{tong2023mass, song2020adversarial}. Studies on model collapse have shown that generative models which scrape large amounts of content available online --- and invariably train on the output of predecessor models --- produce an output distribution that degrades at the tails~\citep{shumailov2024ai}. These findings are further formalized by ~\citep{jiang2026artificial}, who warn of two specific dangers in language models: intra-model repetition (one model consistently generating homogeneous responses) and inter-model homogeneity (multiple models producing similar output). Furthermore, there is evidence that even though an output may be theoretically producible by a model, it may not be reachable through the conditioning mechanisms given to a user \citep{vafa2025s}.

In music generation, many ad hoc, specialized control mechanisms for generative models have been proposed~\citep{wu2024music, novack2024ditto, novack2025ditto, hou2025editing,wang2026latent}. Despite this, there is considerably less literature that explicitly studies \textit{what} is producible in large music models. Recent work in interpretability and steerability for large music models has largely shown that while models like MusicGen do encode explicit musical features, it is challenging to be able to steer the models to produce those actual results \citep{wei2024music, singh2025discovering, zhao2025steering}. While there exist studies that analyze whether generative models \textbf{understand} music in a human-realistic manner ~\citep{zang2025you, morais2026investigating} there is minimal literature that analyzes whether generative models \textbf{produce} human-realistic output. Herein enters our work, the first empirical study of what generative music models \textit{can} output, and a reproducible formula for analyzing and comparing new model generations in this respect as they emerge.

\section{Problem Statement}
\label{sec:problem}

We begin with a pretrained model $\mathcal{M}_m$. We define a prompt generation strategy $s$ (we describe specifics below), and use it to generate $N$ prompts to be mapped to corresponding audio samples with $\mathcal{M}_m$:

\begin{equation}
\begin{aligned}
    t_{si} &\sim T(s,c), \qquad
    y_{msi} \sim \mathcal{M}_m(t_{si})
    \quad \forall i \in \{1,\ldots,N\}
\end{aligned}
\end{equation}

where $T$ is a text generator conditioned on a \textbf{prompt strategy} $s$ and a corpus $c$ of base prompts supplying linguistic features such as vocabulary and syntax.

We then extract interpretable features (which can be continuous, e.g., tempo, or categorical,l e.g., key) from each generated audio sample:
\begin{equation}
	F(y_{msi}) = \{F_1(y_{msi}),F_2(y_{msi}),\dots,F_K(y_{msi})\}
\end{equation}

We produce the empirical feature distribution for model $m$, prompt strategy $s$, and feature $k \in \{1,\dots,K\}$:
\begin{equation}
	P_{ms}^{(k)} = \{ F_k(y_{msi}) \mid 1 \leq i \leq N \}
\end{equation}

Let $D(P, P')$ denote a distance (e.g. Wasserstein) between two feature distributions. Then, we can make comparisons between such distributions. We treat the model training regime as a black box, and we only have high-level details for the training data, model architecture, and training techniques such as the number of training points or whether the model is autoregressive or diffusion-based.

This setup supports operationalizing precise questions about the distribution of generated audio. We ask:
\begin{enumerate}

    \item \textbf{What has a larger expected impact on the generated audio: the model or the prompt strategy?} These prompt strategies were designed to elicit different behaviors from the same model. Thus, if changing the prompt strategy produces smaller changes than changing the model, then the model is the primary determinant of characteristics of its outputs. In contrast, larger between-strategy differences would instead suggest that prompt design yields more substantial control over observed characteristics of the output.

    To see this, let $S = \{s_1,\dots,s_L\}$ be the set of prompt strategies and $\mathbb{M} = \{\mathcal{M}_1,\ldots,\mathcal{M}_M\}$ the set of models. For each feature $k$, we measure:
    \begin{align}
    \label{eq:model-vs-prompt}
	   \mathcal{D}_{\mathrm{model}}^{(k)} &= \left\{D_k\left(P_{ms}^{(k)},P_{m's}^{(k)}\right)\mid s\in S,\ 1\leq m<m'\leq M\right\}\\
       \label{eq:model-vs-prompt-2}
       \mathcal{D}_{\mathrm{prompt}}^{(k)} &= \left\{D_k\left(P_{ms}^{(k)},P_{ms'}^{(k)}\right)\mid m\in \mathbb{M},\ 1\leq s<s'\leq L\right\}
    \end{align}
    For continuous features we use Kolmogorov--Smirnov (KS) distance as $D_k$, and we use total variation (TV) distance for categorical features.
    \item \textbf{To what degree do generated outputs reproduce the range of characteristics found in human-made music?} This is related to concerns about monoculture and homogenization in generative models~\citep{raghavan2024competition,jiang2026artificial}. For this, we examine how far generated outputs extend toward the central limits, tails, and uncommon categories observed in datasets consisting of human-made music.

    For this, we require a human-generated ``reference'' dataset. Let $\mathcal{R}=\{r_1,\dots,r_H\}$ be a collection of human-produced clips. Then, for feature $k$, we have empirical reference distribution:
    \begin{equation}
    \label{eq:real-data}
	   P_{\mathcal{R}}^{(k)} = \left\{F_k(r_i)\mid r_i\in\mathcal{R}\right\}
    \end{equation}
    
    For continuous features, we use \textbf{interval-width ratio} and \textbf{relative probability mass}, e.g. for quantiles $0 \leq a < b \leq 1$, we define the ratio of interval widths using the quantile function $Q$:
    \begin{equation}
    	W_{ms}^{(k)}(a,b) = \frac{Q_b\left(P_{ms}^{(k)}\right) - Q_a\left(P_{ms}^{(k)}\right)}{Q_b\left(P_{\mathcal{R}}^{(k)}\right)-Q_a\left(P_{\mathcal{R}}^{(k)}\right)}
    \end{equation}
    
    For region $A$ with support in the reference distribution, we use relative mass ratio ($\Pr$ is a probability measure):
    \begin{equation}
    	M_{ms}^{(k)}(A) = \frac{\Pr_{x\sim P_{ms}^{(k)}}\left[x\in A\right]}{\Pr_{x\sim P_{\mathcal{R}}^{(k)}}\left[x\in A\right]}
    \end{equation}

    If $A$ corresponds to the distribution tail (e.g. top or bottom 5\%), then this is tail probability mass. For categorical features, we look at category concentration, e.g., examining largest category share or effective number of categories. Finally, we supplement such feature-based comparisons with analysis of joint feature and embedding spaces.
\end{enumerate}

\section{\madlibs}
\label{sec:methods}

\subsection{Prompting Large Music Models}
\subsubsection{Prompt Generation}
\label{subsec:prompting}

Our first task is to generate prompts under a strategy $s$, as introduced in the problem statement in \S\ref{sec:problem}:
\begin{equation}
    t_{si} \sim T(s,c), \qquad \forall_i \in \{1,\dots,N\}
\end{equation}
where $c$ is a source corpus of captions which supplies domain vocabulary and structural statistical properties. Each strategy $s$ thus serves as a modeling layer between this source corpus material (for which we rely on \citet{manco2023song}) and generative models, and induces a prompt distribution $q_s(t \mid c)$; the corresponding audio distribution for model $\mathcal{M}_m$ is:
\begin{equation}
    p_{ms}(y) = \int p_m(y\mid t)q_s(t\mid c)\,\mathrm{d}t
\end{equation}

The downstream empirical feature distribution $P_{ms}^{(k)}$ can then change because of the music model, the prompt strategy, or the interaction between these two variables. The goal of this approach is to produce enough varied conditions to estimate output distributions without restricting the analysis to a manually selected prompt set. The strategies we prototype also progressively introduce musical keywords, corpus-derived statistical properties, and additional heuristics for varying the properties of produced prompt language. Agreement across these different methods would thus help provide evidence of the generality of obtained results.

\begin{enumerate}
	\item \textbf{Blank/constant.} Repeats the same fixed string for every sample (empty prompt by default, and in all of our experiments). This is the lowest-information control in the study, serving as a minimal conditioning baseline. Comparing it with non-empty strategies can help evaluate how much prompt text induces distribution changes altogether
	\item \textbf{Keyword mixing.} This introduces coarse musical semantics. Samples one style label, one mood label, and one instrumentation label and concatenates them into a brief prompt as \textbf{\texttt{\textcolor{blue}{<mood>} \textcolor{red}{<style>} featuring \textcolor{green!60!black}{<instrument>}}}:
    \begin{equation*}
\underset{\color{blue}{\mathrm{mood}}}
  {\textit{\color{blue}{tense}}}
\
\underset{\color{red}{\mathrm{style}}}
  {\textit{\color{red}{lo-fi hip hop}}}
\ \textit{featuring}\
\underset{\color{green!60!black}{\mathrm{instrumentation}}}
  {\textit{\color{green!60!black}{acoustic guitar}}}
\end{equation*}
The 4 moods $\times$ 4 styles  $\times$ 4 instruments produce $4^3 = 64$ possible prompts, which we then sample with replacement.
	\item \textbf{Mad Libs family.} Introduces controlled syntactic structure by separating prompt generation into two components: a structural scaffold and lexical slot fillers. Prompts are constructed from coarse part-of-speech slots (such as noun, verb, adjective, adverb, article, and pronoun), with slot values sampled from the music-domain lexico. We have two variants:
		\begin{enumerate}
			\item The corpus-induced \textit{template} derives prompt templates induced from the caption corpus. We tokenize each caption, replace all recognized adjectives/nouns/verbs/adverbs with ordered part-of-speech fields, and keep function words, punctuation, and unmatched tokens as anchors. We then sample templates and template-specific field values per their counts in the corpus. For simplicity, we keep captions with 5--12 tokens and $\geq 3$ replaceable fields, and then keep the 500 most frequent induced templates.
			\item The (shallow) \textit{PCFG} variant estimates a probabilistic grammar from observed part-of-speech sequence counts, to better match the structural statistics of the original corpus. Sampling from this strategy involves drawing from these skeletons according to their empirical frequency, and then filling parts-of-speech slots from the music-domain lexicon
		\end{enumerate}

        For example:
        \begin{equation*}
    \underset{\color{blue}{\mathrm{Adj}}}
      {\underline{\textit{\color{blue}{slow}}}}
    \ \textit{and}\
    \underset{\color{blue}{\mathrm{Adj}}}
      {\underline{\textit{\color{blue}{arpeggiated}}}}
    \ \textit{solo}\
    \underset{\color{red}{\mathrm{N}}}
      {\underline{\textit{\color{red}{piano}}}}
    \ \textit{with}\
    \underset{\color{red}{\mathrm{N}}}
      {\underline{\textit{\color{red}{reverb}}}}
    \end{equation*}
    
	\item \textbf{Dependency-graph prompts.}
    Imposes stronger grammatical structure by preserving coarse-grained positional relationships between words. It organizes prompts around a heuristic ``root token'' (first token recognized as either noun or verb) and then tags each token's type and position relative to the root as a simple approximation of its role in the sentence. Roles include \texttt{root\_verb}, \texttt{root\_noun}, \texttt{comp\_verb}, \texttt{obj\_noun}, \texttt{pre\_adv}, and \texttt{post\_adv}, and adjectives are classified using the nearest noun. Other tokens are held fixed, and the resulting sequence is then a template, which allows us to roughly approximate sentence grammatical organization without requiring full dependency parsing (since this would introduce another learned model and complicate attribution of downstream differences). We keep the 600 most frequent templates containing 5--14 words and $\geq 3$ replaceable positions. For example:

    \begin{center}
    \begin{minipage}{0.96\linewidth}
    \small
    \textit{Caption:}
    \begin{equation*}
    \underset{\color{green!60!black}{\mathrm{root\ verb}}}
        {\textit{\color{green!60!black}{listen}}}
    \textit{:}\quad
    \underset{\color{red}{\mathrm{object\ noun}}}
        {\textit{\color{red}{piano}}}
    \quad
    \underset{\color{orange!90!black}{\mathrm{complement\ verb}}}
        {\textit{\color{orange!90!black}{drifts}}}
    \quad
    \underset{\color{blue}{\mathrm{post\mbox{-}root\ adverb}}}
        {\textit{\color{blue}{softly}}}
    \quad
    \textit{at}
    \quad
    \underset{\color{red}{\mathrm{second\ object\ noun}}}
        {\textit{\color{red}{night}}}
    \end{equation*}
    
    \textit{Template:}\par
    \begin{center}
    {\ttfamily
    \textcolor{green!60!black}{\{root\_verb\}}:
    \textcolor{red}{\{obj\_noun\}}
    \textcolor{orange!90!black}{\{comp\_verb\}}
    \textcolor{blue}{\{post\_adv\}} at
    \textcolor{red}{\{obj\_noun2\}}
    \par}
    \end{center}
    \end{minipage}
    \end{center}
\end{enumerate}

Note that these strategies are diagnostic heuristics, i.e. they have no claim to optimality, representativeness, or exhaustive coverage, but they provide simple, transparent reference conditions.

\begin{figure*}[!htb]
    \centering
    \includegraphics[width=\linewidth]{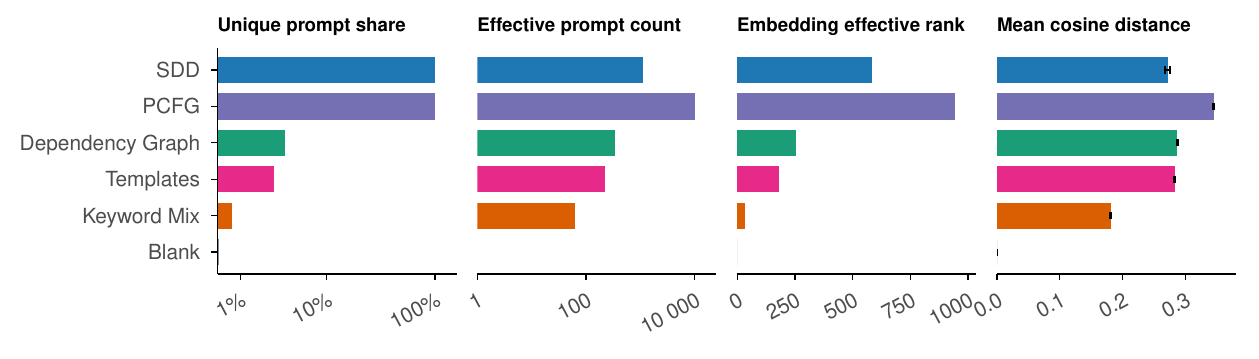}
    \caption{Characterizing information content in generated prompts: unique share (number of unique prompts / total prompts), effective prompt count (number of unique prompts), effective rank of embedding matrix ($\operatorname{erank}(A) = \exp \left( H\left(\frac{\sigma_i(A)}{\sum_j \sigma_j(A)}\right) \right)$, entropy of normalized singular values), and mean pairwise cosine embedding distance.}
    \label{fig:prompts_prompt_support_rank}
\end{figure*}

In \Cref{fig:prompts_prompt_support_rank}, we analyze the distribution of generated prompts. Since we sample \textit{with replacement}, only the PCFG strategy comprises all 10,000 unique prompts. This figure shows that different strategies clearly produce differing levels of diversity and information-richness, which should in principle correspond to differences in the output distribution.

\subsubsection{Generation Models}
\label{subsec:models}

We use three models $\mathcal{M}$ for our experiments: \textbf{\textcolor{MColor}{M}}usicGen \citep{copet2023simple}, \textbf{\textcolor{AColor}{A}}CE-Step \citep{gong2025ace, gong2026ace}, and \textbf{\textcolor{DColor}{D}}iffRhythm \citep{ning2025diffrhythm,jiang2025diffrhythm}. \textbf{\textcolor{MColor}{M}}usicGen is an open-weight autoregressive-only model used in other work exploring the behavior of generative music models \citep{wei2024music, singh2025discovering, zhao2025steering}. We then use \textbf{\textcolor{AColor}{A}}CE-Step and \textbf{\textcolor{DColor}{D}}iffRhythm as baselines for modern diffusion transformers. 

 We choose these models as they are all open-source, open-weight, and are some of the most widely used of the past few years. While we hoped to also include Lyria~\citep{team2025live}, API rate limits did not allow us to test it at scale.

\subsubsection{Reference Dataset}

We use the MTG Jamendo~\citep{bogdanov2019mtg} dataset and the Song Describer Dataset (SDD)~\citep{manco2023song} aa reference distributions, where SDD is a subset of MTG Jamendo audio paired with captions. We need a dataset of human-made music to study \mbox{$P_{\mathcal{R}}^{(k)} = \left\{F_k(r_i)\mid r_i\in\mathcal{R}\right\}$}, an empirical reference distribution of all features. For instance, humans may exhibit mode-seeking behavior on attributes like song key or tempo, which can only be estimated if we have human-made music audio data.

MTG Jamendo contains over 50,000 audio tracks, all with permissive licenses. We acknowledge two key limitations in using MTG Jamendo to approximate human-made music:

\begin{enumerate}
    \item Music in the public domain is typically different than commercial music, and therefore may have different feature distributions. However, we postulate that commercial music has a wider distribution than MTG Jamendo with respect to the music features we extract. Therefore, if a music model has a similar distribution to MTG Jamendo, we assume that the music model has a \textit{narrower} distribution than commercial music and this thus serves as a conservative reference. We make this assumption because all music models included in this study are trained on significantly larger, and partially commercial, music datasets.
    \item MTG Jamendo only has 55,000 audio tracks in total, from which we draw $H=$10,000. Even if the songs in MTG Jamendo were of similar quality to commercial music, our sample is modest and thus likely to yield a conservative feature distribution estimator anyway.
\end{enumerate}

As such, we believe that these limitations do not greatly impair our inferences about distribution coverage and other related properties, but rather may bias them to be even more conservative than reality.

\subsubsection{Generation Strategy}

For each of the prompting strategies (blank, keyword mix, templates, dependency graphs, and PCFG), we sample $N=10,000$ prompts \textit{with replacement} (since some strategies have limited vocabulary). We sample an audio file for each prompt as $y_{msi} \sim \mathcal{M}_m(t_{si}), \forall_i \in \{1,\ldots,N\}$. For ACE-Step and DiffRhythm, we generate 90-second segments. For MusicGen, we generate the maximum length of 30 seconds. We generate only instrumental tracks for consistency across models and to avoid confounding by lyrics. We also use all three models to \textit{regenerate} outputs for each of the 1,106 captions of the Song Describer Dataset used as prompts. We use the regenerated SDD output as a directly comparable first baseline, along with $H=10,000$ randomly sampled songs from MTG Jamendo~\citep{bogdanov2019mtg} (prompt-free) as our large-scale real reference distribution. Since SDD is a subset of MTG Jamendo, we note that our random MTG Jamendo subset contains 208 SDD songs (29.5\% of the SDD dataset).

\subsection{Analysis}\label{sec:experiment}

With the music generated, we now seek to understand it. Here, this means extracting features from the music that are human-interpretable and comparable between models $F(y_{msi}) = \{F_1(y_{msi}),F_2(y_{msi}),\dots,F_K(y_{msi})\}$. Later, we refer to these features as the \textit{musical distribution} of the output. We acknowledge that there are many definitions of the word musical; we call this musical because we are looking at components that are extractable with canonical music information retrieval techniques.

\textbf{Key and Note Detection:} We use SKey~\citep{kong2025s} to detect the keys of generated songs. We define the overall key of the song as the mode of keys in all overlapping 30-second windows. We also capture the relative density of pitch classes with a chromogram sum.

\textbf{Beat Detection:} We use \textit{Beat This!}~\citep{foscarin2024beatthis} library to produce a list of timestamps for each beat and downbeat, which we use to approximate both tempo (here, \textit{downbeats}-per-minute rather than beats-per-minute to account for subdivision-level noise we observed) and time signature (here, beats between consecutive downbeats).

\textbf{Instrument Detection:} We use Music Flamingo \citep{ghosh2026music} to produce instrument tags for each output track. We recompute instrument tags for our MTG Jamendo subset and SDD baselines using Music Flamingo for consistency across distributions. For both reference datasets, we remove the ``vocals'' tag since we are comparing against instrumental generations. For each output distribution, we analyze how many and which instruments appear per track.

\textbf{Dynamics Analysis:} To analyze the dynamics of output audio, we extract two metrics from the root mean squared (RMS) amplitude of the windowed audio signal: (1) the longest continuous silence and (2) the standard deviation of RMS. We calculate (1) by computing RMS for 30 ms windows, with silence defined if $RMS<10^{-3}$. There is no universally agreed threshold for silence in RMS values, but we can adopt a fixed one here since our study compares audio across sources using the same metric throughout. We are interested in the longest silence to capture the presence of musical ideas like breaks or breath. We use (2) the standard deviation of RMS to consider how much the dynamics of the audio vary over time. For example, a crescendo or decrescendo would yield a higher RMS standard deviation relative to audio without that dynamic shift.

\textbf{Prompt Adherence: } To measure prompt-adherence, we compute cosine similarity in the CLAP embedding \citep{wu2023large} space for prompts and their corresponding outputs.

\section{Results}

\begin{figure*}[!htb]
    \centering
    \includegraphics[width=\linewidth]{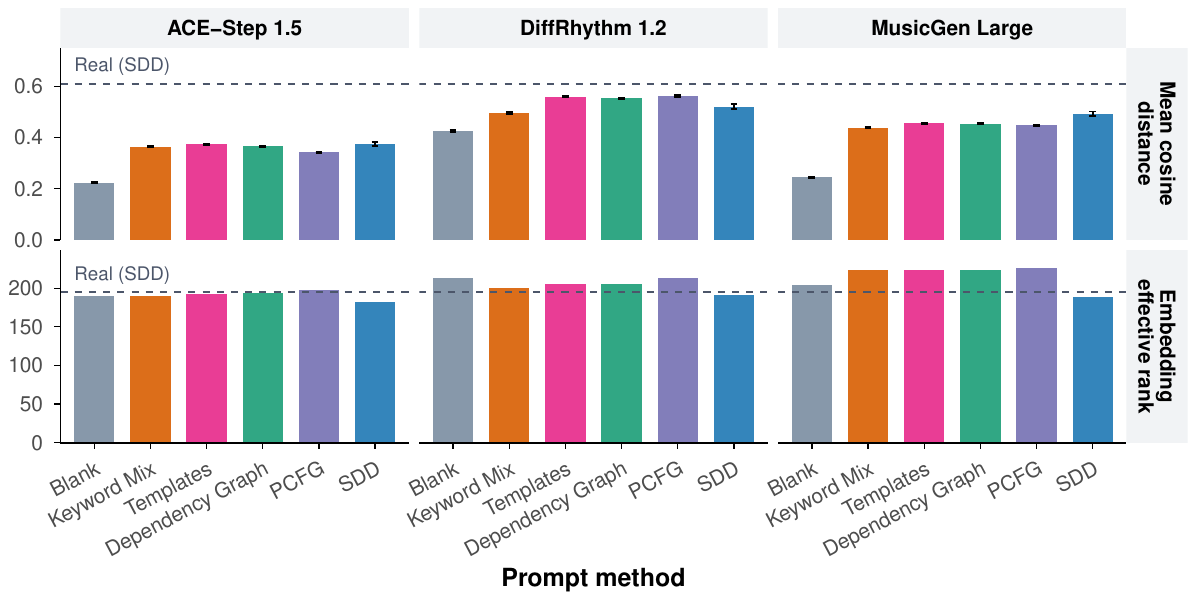}
    \caption{Diversity metrics for CLAP embeddings of audio outputs from each prompting method, model, and (dashed line) real reference SDD audio. Note: for effective rank, the metric is upper bounded by $\min\left(d, N - 1\right)$, and $N$ (number of tracks) is much smaller for SDD ($\approx1k$) and Real (SDD)$\approx700$ than the synthetic datasets ($N=10k)$.}
    \label{fig:prompts_audio_embedding_diversity}
\end{figure*}

\subsection{Model Output is Largely Invariant to Prompt Diversity}
We find the musical output distribution varied minimally with prompt diversity, but was rather dominated by model identity using \Cref{eq:model-vs-prompt,eq:model-vs-prompt-2}. We have five pseudo-random prompt generation strategies with varying degrees of diversity (\Cref{fig:prompts_prompt_support_rank}), which allows us to study how this affects the outputs.

\begin{table}[!t]
\centering
\small
\setlength{\tabcolsep}{4pt}
\begin{minipage}[t]{0.48\linewidth}
    \centering
    \caption{Paired $t$-tests of between-prompt-strategy/between-model distance ratios. Metric is median, mean, or max pairwise KS/TV distance. $p$ tests whether prompt-strategy distances are \textit{smaller} than model distances, all $p<0.02$ after Benjamini-Hochberg correction.}
    \label{tab:distance_ratios}
    \begin{tabular}{@{}clrrr@{}}
    \toprule
    Blank Included & Metric & Med. ratio & Mean-$\Delta$ & $p$-BH \\
    \midrule
    \faTimesCircle & Max & 0.78 & -0.14 & 0.016 \\
    \faTimesCircle & Mean & 0.36 & -0.21 & 0.001 \\
    \faTimesCircle & Median & 0.24 & -0.27 & 0.001 \\
    \midrule
    \faCheckCircle & Max & 0.95 & -0.08 & 0.016 \\
    \faCheckCircle & Mean & 0.49 & -0.19 & 0.001 \\
    \faCheckCircle & Median & 0.36 & -0.24 & 0.001 \\
    \bottomrule
    \end{tabular}
\end{minipage}\hfill
\begin{minipage}[t]{0.48\linewidth}
    \centering
    \caption{Average Longest per-track Silence (AvLS) and probability of any silence occurring in a given track.}
    \label{tab:silence}
    \begin{tabular}{@{}lrr|rr@{}}
    \toprule
    Source & AvLS (ms) & $p$ & $P$(Any Sil.) & $p$ \\
    \midrule
    Jamendo & 135.2 & --- & 0.072 & --- \\
    ACE-Step & 36.9 & $0.06$ & 0.091 & $<.0001$ \\
    DiffRhythm & 309.4 & $0.002$ & 0.084 & $<.0001$ \\
    MusicGen & 0.35 & $0.01$ & 0.003 & $<.0001$ \\
    \bottomrule
    \end{tabular}
\end{minipage}
\end{table}

Given that the prompt set generated by PCFG has significantly greater prompt diversity, we would expect the musical output of a model prompted with it to be significantly more diverse than that of a less diverse prompt corpus. However, we empirically observe that the distribution of the musical features is largely unchanged by this, as observed in \Cref{fig:gen_key_instrument,fig:gen_outcome_distributions,fig:gen_support_intervals}. For example, \Cref{fig:gen_key_instrument} shows detected instruments and key for each model and prompt strategy, which are similar for each model irrespective of prompt strategy. \Cref{tab:distance_ratios} gives statistical evidence of this across features.

We conclude from this that the purported creative affordances of ``careful'' or ``expressive'' prompting may not actually be captured by the models' empirical expressive support. For example, the Keyword Mix strategy had only 64 unique prompts, but the number of songs that had electric guitar-generated or were in the key of G major was essentially the same for more expressive prompters. This could point toward an upper bound on the amount these models are capable of capturing musical variation through prompting. Notably, the blank prompt strategy is an extreme edge case where every prompt is the same (``''), and the output distribution is thus different than other prompt strategies for some models, most clearly with MusicGen.

\subsection{Each Model Has Learned Idiosyncratic Properties}

We also observe that distribution differences between models persist under prompt variation (which \Cref{tab:distance_ratios} also supports). For example, ACE-Step 1.5 is not likely to generate synthesizer or electric guitar timbres, but those timbres are common in generated output from DiffRhythm 1.2 and MusicGen Large (see \Cref{fig:gen_key_instrument}). Rather, each model exhibits a recognizable default musical regime, i.e. a region of tempo, instrumentation, harmonic profile, dynamic variation, and other attributes that it repeatedly returns to across prompt distributions.

\subsection{Analysis of Musical Characteristics}

\begin{figure}[!t]
    \centering
    \begin{minipage}[t]{0.45\linewidth}
        \vspace{0pt}\centering
        \includegraphics[width=\linewidth]{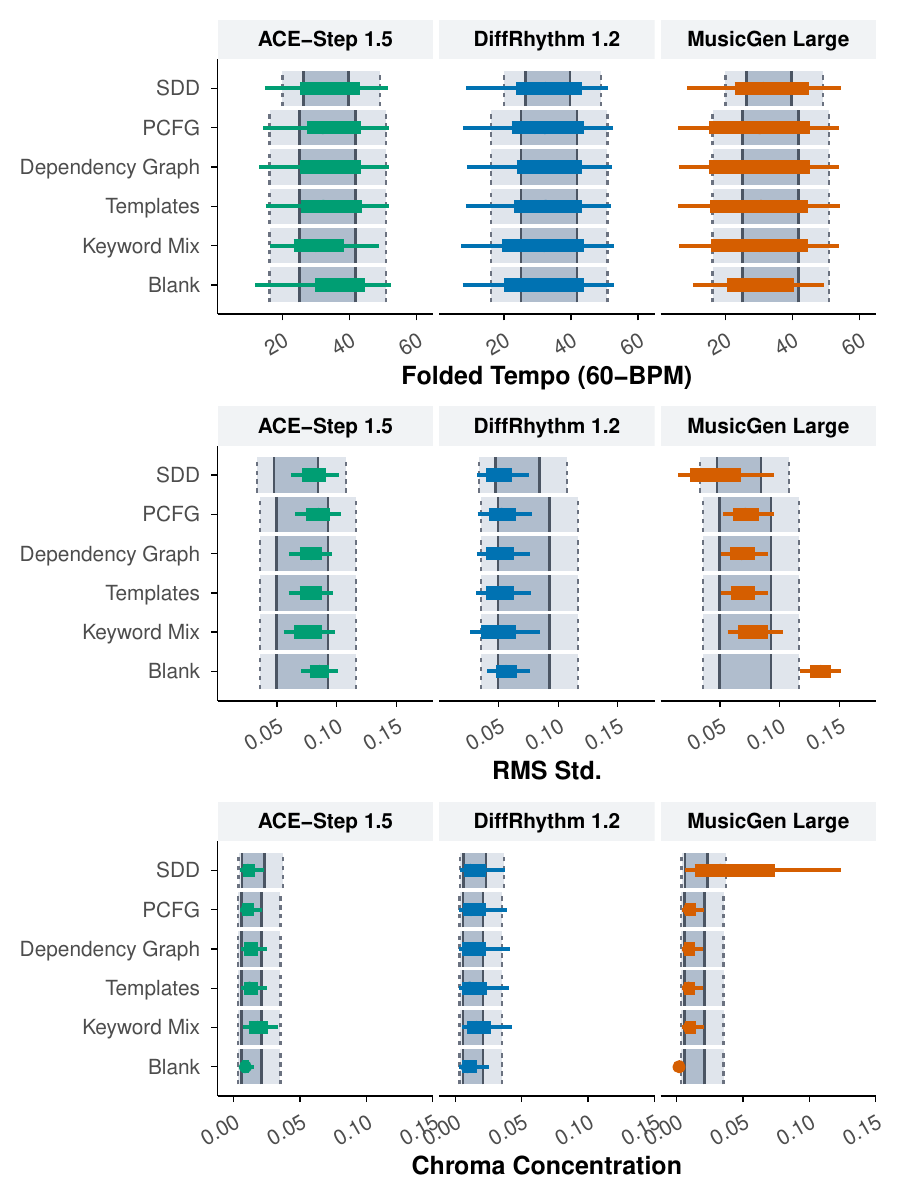}
        \caption{Coverage of tempo (folded, \% 60), dynamic variation (RMS Std.) and chroma concentration (1 - normalized chroma entropy). Thick and thin regions of lines show IQR (25th/75th \%-ile) and 90th/10th \%-iles, respectively. Inner and outer gray bands show the same quantiles for the real (SDD top, Jamendo otherwise) distribution.}
        \label{fig:gen_support_intervals}
    \end{minipage}\hfill
    \begin{minipage}[t]{0.51\linewidth}
        \vspace{0pt}\centering
        \includegraphics[width=\linewidth]{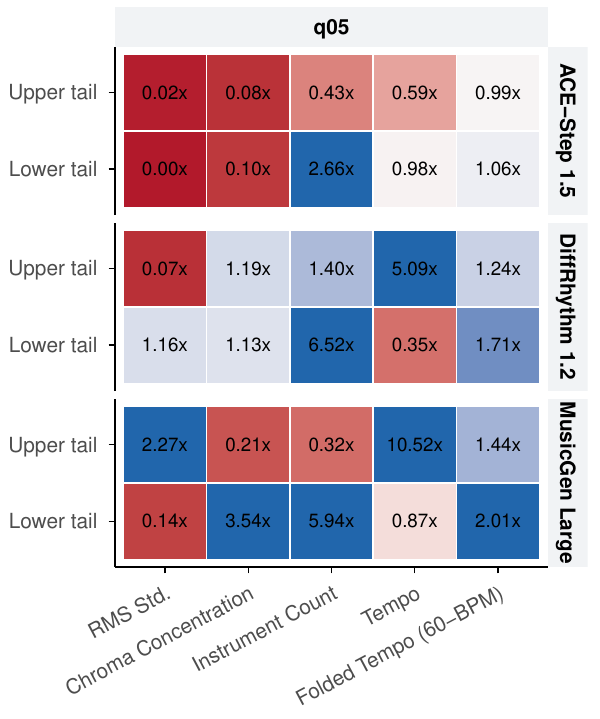}
        \caption{Lower- and upper-tail (5th/95th \%-ile) ratios relative to Real (SDD). A value of 1 matches the reference tail mass, values above 1 overrepresent extremes, and values below 1 underrepresent them.}
        \label{fig:gen_tail_reach}
    \end{minipage}
\end{figure}

\Cref{fig:gen_support_intervals} covers an analysis of extracted folded tempo (estimated tempo \% 60), RMS std., and chroma concentration. If the \textbf{folded tempo} distribution is wider, the corpus likely has more tempo variation. In this respect, models match the real distribution or are even flatter (e.g. MusicGen). We interpret \textbf{RMS std.} as a signal of dynamic variation, and its distribution is how much this varies across tracks (e.g. via stylistic differences). Variation could be due to macro crescendo/decrescendo dynamic shift, or a more granular variation like a heavily syncopated beat. We find that across all three models and all prompt strategies, the RMS std. was less varied than the baseline MTG-Jamendo dataset. We interpret this as implying that the dynamics of generated music are less context-dependent and more uniform. We also observe that MTG-Jamendo has some music that varies dynamically more than \textit{any} generated music, other than blank prompts into MusicGen Large. \textbf{Chroma concentration} is inverse summed-chromagram entropy. A chroma concentration closer to zero means a flatter distribution, i.e. all pitch classes are equally present in the aggregated output. Here, DiffRhythm output is similar to Jamendo, but ACE-Step 1.5 and MusicGen Large are shifted left and narrow, meaning that they do not generally prefer certain pitches compared to MTG-Jamendo.

\begin{figure}[t!]
    \centering
    \includegraphics[width=0.95\linewidth]{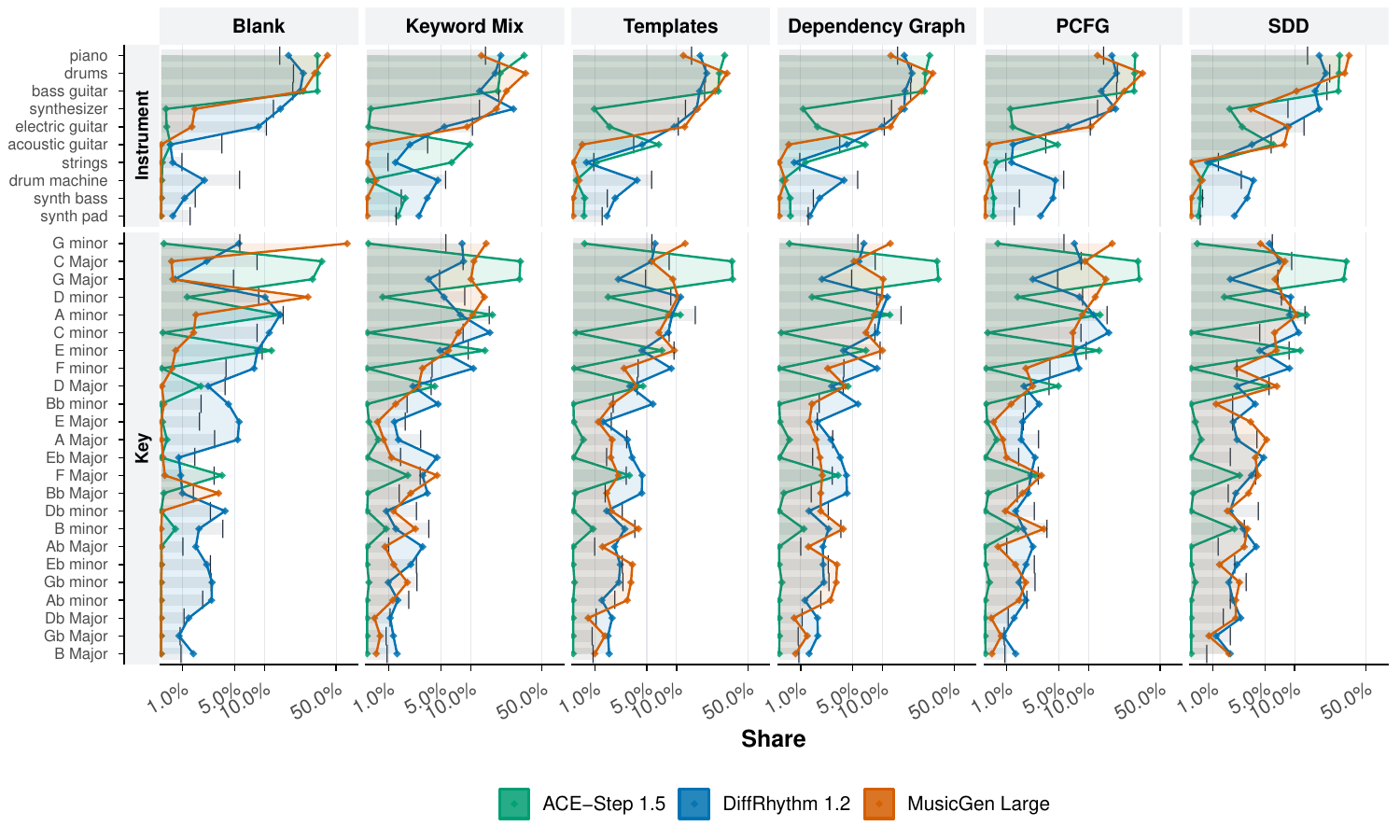}
    \caption{Instrument and key representations relative to the Jamendo baseline (shown as \textbf{--}). Both distributions show greater inequality than the reference distribution, with relative prompt strategy invariance.}
    \label{fig:gen_key_instrument}
\end{figure}

\begin{figure}[b!]
    \centering
    \includegraphics[width=0.95\linewidth]{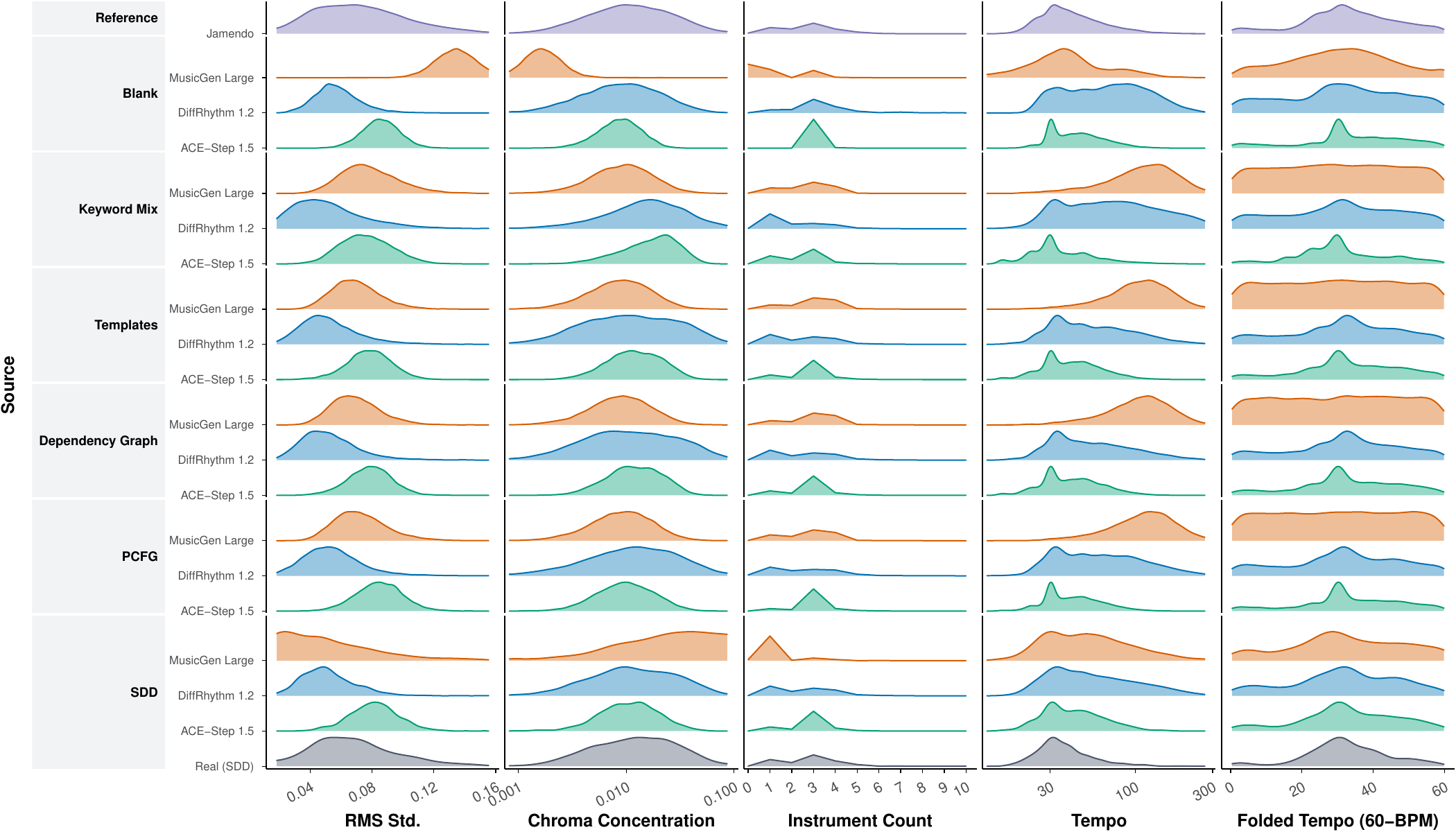}
    \caption{Density/ridge plots showing probability mass of various features for all real and synthetic distributions.}
    \label{fig:gen_outcome_distributions}
\end{figure}

\Cref{fig:gen_tail_reach} shows the tail mass for these variables, and also for \textbf{instrument counts}. For all three models, the lower tail is overrepresented compared to the SDD dataset by a significant margin (over 2.5$\times$ for all). Additionally, for two of the three models, the upper tail for instrument count is significantly \textit{under}represented. All three models are much more likely to generate songs with fewer instruments than the reference, and the distribution of instrument count is narrower, as shown in \Cref{fig:gen_outcome_distributions}. \Cref{fig:gen_key_instrument} also shows that piano, drums, and bass guitar comprise a larger share of generated music than Jamendo. We interpret this as generative models overrepresenting common instruments.

\Cref{fig:gen_key_instrument} offers evidence that models tend to overrepresent common keys such as C and G major. ACE-Step even largely fails to produce songs for certain keys regardless of the prompting strategy. Our interpretation is that generative models tend to overrepresent more common keys, which may be more abundant in online training data, while collapsing (or outright excluding) the tails.

Next, we investigate the models' capacity for producing silence. John Cage said ``try as we may to make a silence, we cannot.'' Can generative models? In \Cref{tab:silence}, we find that 2/3 models (ACE-Step and DiffRhythm) are more likely to produce silence than Jamendo, whereas MusicGen produces almost none. DiffRhythm also produces longer silences on average.

\begin{table}[!t]
\centering
\small
\begin{minipage}[t]{0.48\linewidth}
    \centering
    \caption{CLAP similarity estimated marginal means (EMMs) after linear regression with cluster-robust SEs. $p$ vs. max is the Benjamini-Hochberg-adjusted contrast against the generator with the largest EMM.}
    \label{tab:clap_stats}
    \begin{tabular}{@{}lrr@{}}
    \toprule
    Source & EMM [95\% CI] & $p$ vs. max \\
    \midrule
    ACE-Step & 0.245 [0.243, 0.246] & --- \\
    DiffRhythm & 0.216 [0.215, 0.218] & $<.0001$ \\
    MusicGen & 0.163 [0.162, 0.164] & $<.0001$ \\
    \bottomrule
    \end{tabular}
\end{minipage}\hfill
\begin{minipage}[t]{0.48\linewidth}
    \centering
    \caption{Audio embedding cosine-distance diversity aggregated across prompt methods. Values are mean cosine distance [95\% CI]; $p$ vs.\ SDD is an approximate BH-adjusted contrast using the reported confidence intervals.}
    \label{tab:embedding_variety}
    \begin{tabular}{@{}lrr@{}}
    \toprule
    Source & Pairwise dist. [95\% CI] & $p$ vs.\ SDD \\
    \midrule
    SDD & 0.61 [0.60, 0.62] & --- \\
    ACE-Step & 0.33 [0.33, 0.33] & $<.0001$ \\
    DiffRhythm & 0.52 [0.52, 0.53] & $<.0001$ \\
    MusicGen & 0.42 [0.42, 0.42] & $<.0001$ \\
    \bottomrule
    \end{tabular}
\end{minipage}
\end{table}

Finally, we look at \textbf{CLAP embeddings of the generated audio}, which are less interpretable but may capture subtle relationships not addressable via known metrics. First, we consider \textbf{prompt adherence} (\Cref{tab:clap_stats}). We find that ACE-Step has the highest CLAP-rated prompt adherence, then DiffRhythm, then MusicGen. Next, we look at \textbf{output diversity} (\Cref{fig:prompts_audio_embedding_diversity}) and find that all generated model sets are \textit{less} diverse than SDD, $p<.0001$ for all, despite containing over 10$\times$ the count of tracks. Indeed, ACE-Step is the \textit{newest} model, has the \textit{highest} (local) prompt adherence, and the \textit{lowest} output diversity, a troubling trend toward more compressed distributions.

\subsection{Limitations}

Ideally, we would compare the generated output of these music models to a definitive distribution of all human-made music. If such a distribution existed, we would be able to make strong claims about where these text-to-music models fall short of human creative decisions. However, no such data exists (esp. under a permissive license).

For each model, we used default generation parameters, as a typical user would. Since we do not adjust parameters like temperature, top-k/top-p, diffusion sampling steps, etc., we do not analyze their impact on the output. The consistency in comparisons is also limited by the maximum 30-second length of MusicGen, and we acknowledge that this may impact the downstream analyses and comparisons across models.

\section{Conclusion}
Evaluations of music generation models arguably have a missing-the-forest-for-the-trees problem. In focusing our attention on the quality of generated outputs, we may miss important properties of their distribution, and thus fail to anticipate downstream effects on musical culture more broadly. In this work, we offer a first glance at this forest, this broader distribution, in the hope of motivating more analysis of how generative outputs might behave at the population level and how we might steer such effects to be commensurate with what we desire.

For instance, our findings demonstrate that if a human prompter wants to exert the most creative influence over their final output, the model they choose to use could affect the musical qualities of their output greatly, beyond just their prompting. We speculate that this might create a kind of agency inversion, wherein the user appears to exercise control through natural language, but at the population level the model's learned distribution exerts stronger control over the musical outcome than the user's efforts. In the worst case, this runs the risk that, absent any principled interventions, generative music models slowly normalize a narrower musical world while still giving users the feeling of choice.

We therefore urge the community to build upon this analysis to discover \textbf{why} these behaviors occur across models and explore model design methodologies in terms of their effects on population-level pluralism and not just output-level quality.

\section{Reproducibility}

All code for \madlibs can be found at \hyperref[https://anonymous.4open.science/r/MADlibs-CBD8/]{https://anonymous.4open.science/r/MADlibs-CBD8/}

We provide code examples for the following components of the \madlibs pipeline below.

\paragraph{Prompt Generation}
Generating $N=10,000$ prompts with the dependency graph prompt strategy can be done as follows

\begin{lstlisting}[style=terminal]
python3 run_prompts.py \
    prompt_strategy=dependency_graph \
    run.n_prompts=10000 \
    run.output_dir=output_dir \
    hydra.run.dir=run_dir
\end{lstlisting}

\paragraph{Music Generation}
Once we have our 10,000 prompts, we can generate corresponding music using one of our given models. For this example, we choose ACE-Step:

\begin{lstlisting}[style=terminal]
python3 run_generation.py \
    generator=acestep \
    prompt_strategy=dependency_graph \
    run.output_dir=output_dir \
    hydra.run.dir=run_dir
\end{lstlisting}

\lstset{
    language=Bash,
    basicstyle=\ttfamily\small,
    keywordstyle=\color{blue},
    commentstyle=\color{gray},
    stringstyle=\color{red},
    numbers=left,
    numberstyle=\tiny,
    frame=single,
    breaklines=true,
}

\paragraph{\madlibs Analysis}
After we have generated the corresponding 10,000 music clips, we can perform \madlibs analysis. We separate this into two categories: \lstinline{analysis_type=mir} is for any MIR analysis that only requires CPU (e.g. RMS) and \lstinline{analysis_type=gpu} is for any MIR analysis that requires GPU (e.g. instrument tagging). An example of running the MIR analysis pipeline is shown below. 

\begin{lstlisting}[style=terminal]
python3 run_mir.py \
    analysis_type=mir \
    generator=acestep \
    prompt_strategy=dependency_graph \
    run.output_dir=output_dir \
    hydra.run.dir=run_dir
\end{lstlisting}

\section{Ethics and consent}

We did not perform any human studies in this work.

\bibliographystyle{unsrtnat}
\bibliography{references}  

\end{document}